\documentclass{article}

\usepackage[numbers]{natbib}         
\usepackage[colorlinks]{hyperref}    
\usepackage[english]{babel} 
\usepackage{amssymb}
\usepackage{amsmath}
\usepackage{txfonts}
\usepackage{mathdots}
\usepackage[classicReIm]{kpfonts}
\usepackage[dvips]{graphicx} 
\usepackage[a4paper, portrait, margin=1in]{geometry}
\usepackage{tabularx}
\usepackage{longtable}
\usepackage{multirow}
\usepackage{booktabs}
\usepackage[labelsep=period]{caption}
\usepackage{makecell}
\usepackage{arydshln}
\begin{document}
	\setlength{\parindent}{0pt}
	\setlength{\parskip}{1ex}
	
	\textbf{\Large Automatic Treatment Planning using Reinforcement Learning for High-dose-rate Prostate Brachytherapy}
	
	\bigbreak

	Tonghe Wang$^{1*}$, Yining Feng$^{1}$, Xiaofeng Yang$^{2,3,4}$

	1Department of Medical Physics, Memorial Sloan Kettering Cancer Center, New York, NY, 10065
	
	2Department of Radiation Oncology and Winship Cancer Institute, Emory University, Atlanta, GA 30322
	
	3Wallace H. Coulter Department of Biomedical Engineering, Georgia Institute of Technology and Emory University School of Medicine, Atlanta, GA, 30332
	
	4Department of Biomedical Informatics, Emory University, Atlanta, GA 30322

	\bigbreak
	\bigbreak
	\bigbreak

	\textbf{*Corresponding author: }
	
	Tonghe Wang, PhD
	
	Department of Medical Physics
	
	Memorial Sloan Kettering Cancer Center 
	
	1275 York Avenue
	
	New York, NY 10065
	
	E-mail: wangt8@mskcc.org

	\bigbreak
	\bigbreak
	\bigbreak
	\bigbreak
	\bigbreak
	\bigbreak

	\textbf{Abstract}

	Purpose: In high-dose-rate (HDR) prostate brachytherapy procedures, the pattern of needle placement solely relies on physician experience. We investigated the feasibility of using reinforcement learning (RL) to provide needle positions and dwell times based on patient anatomy during pre-planning stage. This approach would reduce procedure time and ensure consistent plan quality.
	
	Materials and Methods: We train a RL agent to adjust the position of one selected needle and all the dwell times on it to maximize a pre-defined reward function after observing the environment. After adjusting, the RL agent then moves on to the next needle, until all needles are adjusted. Multiple rounds are played by the agent until the maximum number of rounds is reached. Plan data from 11 prostate HDR boost patients (1 for training, and 10 for testing) treated in our clinic were included in this study. The dosimetric metrics and the number of used needles of RL plan were compared to those of the clinical results (ground truth).
	
	Results: On average, RL plans and clinical plans have very similar prostate coverage (Prostate V100) and Rectum D2cc (no statistical significance), while RL plans have less prostate hotspot (Prostate V150) and Urethra D20\% plans with statistical significance. Moreover, RL plans use 2 less needles than clinical plan on average.
	
	Conclusion: We present the first study demonstrating the feasibility of using reinforcement learning to autonomously generate clinically practical HDR prostate brachytherapy plans. This RL-based method achieved equal or improved plan quality compared to conventional clinical approaches while requiring fewer needles. With minimal data requirements and strong generalizability, this approach has substantial potential to standardize brachytherapy planning, reduce clinical variability, and enhance patient outcomes.

	
	\bigbreak
	\bigbreak

	\noindent 
	\section{INTRODUCTION}
	
	High-dose-rate (HDR) brachytherapy is one of the most widely used treatment modalities for prostate cancer. During the procedure, typically eight to eighteen needles are interstitially implanted into the prostate under transrectal ultrasound (TRUS) guidance in the operating or procedure room (OR), following patient anesthesia. After implantation, planning images—whose modality may vary depending on institutional protocols—are acquired to facilitate the delineation of target structures and organs-at-risk (OARs), as well as needle digitization within the treatment planning system. These steps are followed by inverse planning optimization of source dwell times.
	
	Although modern treatment planning systems allow for the optimization of dwell times and locations, the initial needle placement remains a critical determinant of the achievable plan quality.\cite{RN1} Currently, needle implantation relies heavily on physician experience and intuition, with no automated guidance for optimal positioning. A suboptimal needle configuration often remains undetected until the planning stage, at which point remedial action—such as the insertion of additional needles—may be required. This is feasible only if the patient is still under anesthesia, but doing so incurs additional OR time, the need for re-imaging, and repetition of the planning process. If anesthesia has already been reversed, re-adjusting needle positions under TRUS guidance is generally impractical. To mitigate this risk, some institutions implant more needles than strictly necessary, which increases procedure time, complicates the digitization process, and may elevate patient risk due to the additional punctures.
	
	These limitations underscore the need for an automated planning system capable of providing physicians with optimized, clinically achievable needle configurations and corresponding dwell time distributions prior to needle insertion. Such a system would offer procedural guidance, reduce the likelihood of replanning, enhance treatment efficiency, and improve patient safety and overall plan quality.
	
	In this study, we propose a reinforcement learning (RL)-based automatic treatment planning framework that generates a complete HDR brachytherapy plan—including needle positions, dwell positions, and dwell times—based solely on segmented targets and OARs. The RL agent iteratively adjusts the placement and dwell times of individual needles using a predefined reward function, and it is trained on a small dataset without requiring ground-truth plans. The output provides physicians with a clinically practical treatment plan prior to needle insertion. To our knowledge, this is the first study to apply reinforcement learning to generate a full treatment plan for prostate HDR brachytherapy in advance of needle implantation.

	\noindent 
	\section{Methods and materials}
	
	\subsection{Data}
	
	In this study, we retrospectively collected a cohort of 11 patients who underwent HDR prostate boost brachytherapy at our institution, each receiving a total dose of 15 Gy. For every patient, planning transrectal ultrasound (TRUS) images were acquired intraoperatively in the operating room (OR). The prostate, rectum, and urethra were contoured by radiation oncologists; catheter digitization was performed by medical physicists; and treatment plans were generated and optimized by physicists and subsequently reviewed and approved by physicians. All planning was conducted using the Vitesse treatment planning system (Varian Medical Systems).
	
	TRUS images were acquired in axial slices and stored in DICOM format with a resolution of 1024 × 768 pixels, a pixel spacing of 0.1 × 0.1 mm², and a slice thickness of 1 mm. Associated RT Plan and RT Structure files were extracted to retrieve catheter positions and organ contours for alignment with the TRUS images.
	For model development, one patient (designated as P01) was randomly selected for training. The remaining ten patients (P02–P11) were used exclusively for evaluation, allowing assessment of the model’s generalizability to unseen data.
	
	\subsection{Reinforcement Learning}
	
	The proposed reinforcement learning (RL) framework is designed to iteratively optimize both the position of each needle and its associated dwell time distribution to maximize a predefined reward function based on clinically relevant dosimetric criteria. The agent sequentially adjusts one needle at a time, modifying both its spatial coordinates and dwell time allocation. Once all needles have been adjusted, the process is referred to as a full round. The agent continues to iterate through multiple rounds until a specified maximum number of rounds is reached. A flowchart outlining the RL workflow for prostate HDR brachytherapy planning is provided in Figure 1. 
	
	\begin{figure}
		\centering
		\noindent \includegraphics*[width=8in, height=6in, keepaspectratio=true]{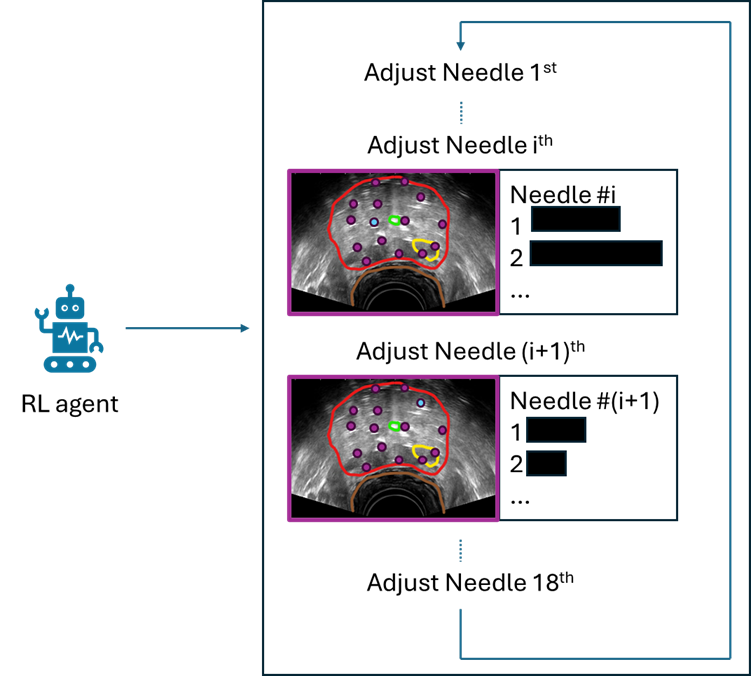}
		
		\noindent Figure 1: Iterative needle adjustment process driven by the RL agent for prostate HDR brachytherapy treatment planning
	\end{figure}
	
	Fig.2 shows the training loop of RL agent. At each time step within a round, the RL agent receives an observation from the environment, denoted as the current state $S_t$. This observation includes anatomical information (e.g., contours of the prostate and organs-at-risk), the current configuration of needles, and dose distribution metrics. The agent extracts features from $S_t$ and passes them through a policy network—in this study, implemented using Proximal Policy Optimization (PPO)—to determine the next action $A_t$, which may involve translating the needle or adjusting dwell times.
	
	Upon executing the action $A_t$, the environment transitions to a new state $S_{t+1}$ and computes an immediate reward $R_{t+1}$ using a reward function that encodes desirable planning objectives (e.g., high target coverage, low OAR dose). During training, the new state and reward are fed back into the agent to update the policy network. The agent learns to increase the likelihood of selecting actions that result in higher cumulative rewards over time.
	
	\begin{figure}
		\centering
		\noindent \includegraphics*[width=6in, height=4in, keepaspectratio=true]{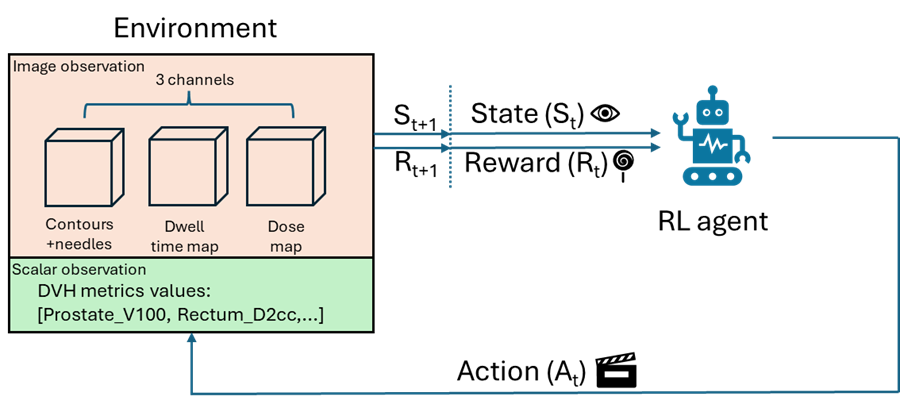}
		
		\noindent Figure 2: Overview of the RL agent training loop
	\end{figure}
	
	In summary, the RL agent continuously refines its policy through repeated interaction with the simulated planning environment, aiming to autonomously generate clinically acceptable HDR treatment plans. The specific components of the environment, action space, reward function, and training protocol are described in the following subsections.
	
	\subsubsection{Environment and Observation Space}
	
	A custom simulation environment was developed to model the needle insertion and treatment planning process for prostate HDR brachytherapy. The simulation space is represented as an 80 × 80 × 80 voxel grid, with an isotropic voxel size of 1 mm³. Patient anatomical structures—namely the prostate, rectum, and urethra—are converted into binary masks and embedded within this 3D environment. Each anatomical mask is assigned a distinct integer value to encode organ type. A total of 18 needles are modeled, all inserted perpendicularly to the axial plane. As a result, the position of each needle can be represented as a 2D coordinate on the axial slice. Needle placement is constrained to lie within the maximum axial projection of the prostate. Each needle contains 16 dwell positions spaced at 5 mm intervals along the superior-inferior axis. Only dwell positions located within the prostate contribute to dose calculations. 
	
	The input to the reinforcement learning agent, referred to as the observation space, consists of both volumetric and scalar components: 
	
	Volumetric input: A three-channel 3D volume of shape 3 × 80 × 80 × 80.
	
	•	Channel 1: A piecewise-constant map encoding the anatomical masks (prostate, rectum, urethra) and needle locations. Needles are visualized as 3 × 3 voxel boxes, with the currently controlled needle differentiated by a unique label value.
	
	•	Channel 2: A map representing dwell times at each dwell position. Dwell positions are encoded at their corresponding 3D coordinates, with voxel intensity proportional to the dwell time in seconds.
	
	•	Channel 3: The cumulative dose distribution map generated from the current needle configuration and dwell time settings.
	
	Scalar input: A list of selected dose-volume histogram (DVH) metrics used to summarize the plan quality:
	
	•	Prostate V100 (volume receiving 100\% of the prescription dose)
	
	•	Prostate V150
	
	•	Rectum D2cc
	
	•	Urethra D20%
	
	•	Non-Prostate Rx Dose Volume (\%), defined as the ratio between the volume receiving the prescription dose outside the prostate and the prostate volume
	
	The environment is initialized with all 18 needles uniformly distributed within the prostate. Initial needle positions are determined by applying a K-means clustering algorithm (k = 18) to the prostate’s largest axial cross-section. The centroid of each cluster is used as the corresponding needle location. Initial dwell times are heuristically assigned based on the spatial proximity to the urethra: for each dwell position, the Euclidean distance to the urethra mask is computed (in voxels), and the dwell time is set to 0.2 times the distance (in seconds), promoting urethral sparing in the initial configuration.
	
	\subsubsection{Network Architecture}
	
	The reinforcement learning agent was implemented using the Proximal Policy Optimization (PPO) algorithm, a widely used policy-gradient method known for its stability and sample efficiency. The input observation space comprises both volumetric image data and scalar features, which are processed through a custom feature extraction module.
	
	The image component of the observation (a 3 × 80 × 80 × 80 volume) is processed using a convolutional neural network (CNN) designed to capture spatial features relevant to anatomy, needle placement, dwell times, and dose distribution. The scalar features, consisting of selected dose-volume histogram (DVH) metrics, are flattened into a one-dimensional vector. The outputs of the CNN and the scalar branch are then concatenated into a single feature vector that represents the current environment state.
	
	This fused feature vector is passed to the PPO policy network, which consists of two fully connected layers with 128 units each and ReLU activation functions. The policy network outputs a continuous action vector that specifies either the adjustment to the needle position or the modification of dwell times for the currently selected needle.
	The network is trained to maximize the expected cumulative reward using PPO's clipped objective function, ensuring that updates to the policy remain within a stable range to prevent performance degradation.
	
	\subsubsection{Action Space}
	
	The action space defines the set of actions available to the reinforcement learning (RL) agent at each decision step. In this study, the RL agent is permitted to adjust both the spatial location and dwell time distribution of a single needle during each iteration.
	
	Specifically, the agent can modify:
	
	•	The 2D coordinates (X, Y) of the needle tip on the axial plane, constrained to lie within the maximum axial projection of the prostate.
	
	•	The dwell times for all 16 dwell positions along the selected needle, with each dwell time bounded within the range of [0, 20] seconds. This upper limit corresponds to the maximum expected dwell time under a 10 Ci source activity assumption.
	
	In total, the action space per needle comprises 18 continuous values: 2 for needle position and 16 for dwell time settings. At each time step, the agent outputs a continuous action vector of length 18 to update the current needle configuration. The modified configuration is then evaluated within the environment to assess its impact on the overall treatment plan.
	
	\subsubsection{Reward Function}
	
	The reward function defines the objective that guides the RL agent during both the training and inference phases. In this study, the reward function was carefully designed to balance multiple clinical priorities in HDR prostate brachytherapy, including target coverage $R_{coverage}$, organ-at-risk (OAR) sparing $R_{constraint}$, and minimization of unnecessary dose deposition outside the target volume $R_{RxVolume}$.
	The total reward $R_{total}$ at each step is composed of three components:
	
	\begin{equation}
		R_{total}=(R_{coverage}+R_{constraint}+R_{RxVolume})/100
	\end{equation}
	Each reward components are defined as follows::
	\begin{equation}
		R_{\text{coverage}} = 
		\begin{cases}
			( {ProstateV100} - G_{{ProstateV100}} ) \times 5 & \text{if } {ProstateV100} < G_{{ProstateV100}} \\
			{ProstateV100} - G_{{ProstateV100}} & \text{if } {ProstateV100} \geq G_{{ProstateV100}}
		\end{cases}
	\end{equation}
	\begin{equation}
		R_{coverage}=-\max(ProstateV150-G_{{ProstateV150}},0)-\max(RectumD2cc-G_{{RectumD2cc}},0)-\max(UrethraD20-G_{{UrethraD20}},0)
	\end{equation}
	where G represents the percentage value of clinical goals as list in Table 1.
	\begin{equation}
		R_{RxVolume}=-\max(V_{RxDose}-G_{RxDose},0)
	\end{equation}
	where $V_{RxDose}$ represents the volume of prescribed dose outside the target divided by the target volume in percentage, and $G_{RxDose}$ is customly defined as 50, i.e. we prefer the volume of prescribed dose outside the target is less than half of the target volume. 
	
	\begin{table}[]
		\caption{Clinical dose goals used in reward function}
		\label{tab:my-table}
		\resizebox{\columnwidth}{!}{%
			\begin{tabular}{lll}
				\hline
				Prostate V100 & \textgreater{}95\% &  \\ \hline
				Prostate V150 & \textless{}50\% &  \\ \hline
				Rectum Dcc & \textless{}70\% &  \\ \hline
				Urethra D20 & \textless{}110\% &  \\ \hline
			\end{tabular}%
		}
	\end{table}

	\subsubsection{Postprocessing}
	
	As the RL agent does not explicitly enforce spatial constraints between needles during training, the generated treatment plans may occasionally contain needles positioned unrealistically close to one another, which is not feasible in clinical practice. To address this, a postprocessing step is applied to merge needles that are spatially clustered too closely.
	
	We use the Density-Based Spatial Clustering of Applications with Noise (DBSCAN) algorithm to identify clusters of needle positions based on their 2D axial-plane coordinates. A maximum distance threshold ($\epsilon$ = 5 mm) is defined, such that any two needles within this distance are considered part of the same cluster.
	
	For each identified cluster, the constituent needles are merged into a single representative needle. The new needle position is computed as the dwell-time-weighted centroid of all needles in the cluster. The updated dwell times are calculated as the element-wise sum of dwell times from the individual needles at each of the 16 dwell positions. This approach preserves both spatial dose distribution and total dose contribution while ensuring a clinically acceptable configuration.
	
	This postprocessing step ensures the final treatment plan adheres to practical implantation constraints, improving its clinical feasibility without requiring retraining of the RL agent.
	
	\subsubsection{Implementation and evaluation}
	The proposed reinforcement learning (RL) framework was implemented using the Stable Baselines3 library, built in Python 3.6 with the PyTorch backend. All training and testing were conducted on a workstation equipped with an NVIDIA A40 GPU with 48 GB of memory.
	
	For each needle, the agent was allowed to propose 32 candidate actions (i.e., different configurations of needle position and dwell times). The environment evaluated the reward for each candidate, and the action yielding the highest reward was selected for execution. The agent sequentially adjusted each of the 18 needles in order, completing one round. After the final (18th) needle, the agent returned to the first needle to begin the next round. A total of 20 rounds constituted one game.
	
	During training, the discount factor $\gamma$ was set to 0.95 and the value function coefficient to 0.4. All other PPO hyperparameters were kept at their default values. The agent was trained for a total of 25 games, which required approximately 5 hours of computation. The resulting trained models were compact, with sizes under 100 MB. During inference, the time required to generate a complete treatment plan for a single patient was approximately 5 minutes.
	
	To quantitatively evaluate the performance of the proposed framework, we compared the RL-generated plans with ground-truth clinical plans using established dose-volume histogram (DVH) metrics. Specifically, we assessed: Prostate V100 and V150, Rectum D2ccand Urethra D20\%. Additionally, we compared the number of needles used in the RL-generated plans versus the clinical plans, to assess the potential for reducing invasiveness and procedure complexity.

	\section{Results}
	Figure 3 presents representative results for two patients, comparing clinical treatment plans with those generated by the proposed reinforcement learning (RL) framework. In both cases, the RL-generated plans utilized fewer needles than the corresponding clinical plans, particularly by reducing the number of centrally placed (internal) needles. Despite the reduced number of needles, the RL plans achieved comparable prostate coverage, as shown by the isodose lines, while demonstrating improved sparing of the urethra.
	
	The dose-volume histogram (DVH) comparisons further support these observations. The RL-generated prostate DVH curves exhibit a sharper decline beyond the prescription dose compared to clinical plans, indicating reduced hotspot regions and a more uniform dose distribution within the target volume. Additionally, RL plans showed lower urethral dose at the D20\% level and reduced low-dose exposure to the rectum.
	
	\begin{figure}
		\centering
		\noindent \includegraphics*[width=6.50in, height=4.20in, keepaspectratio=true]{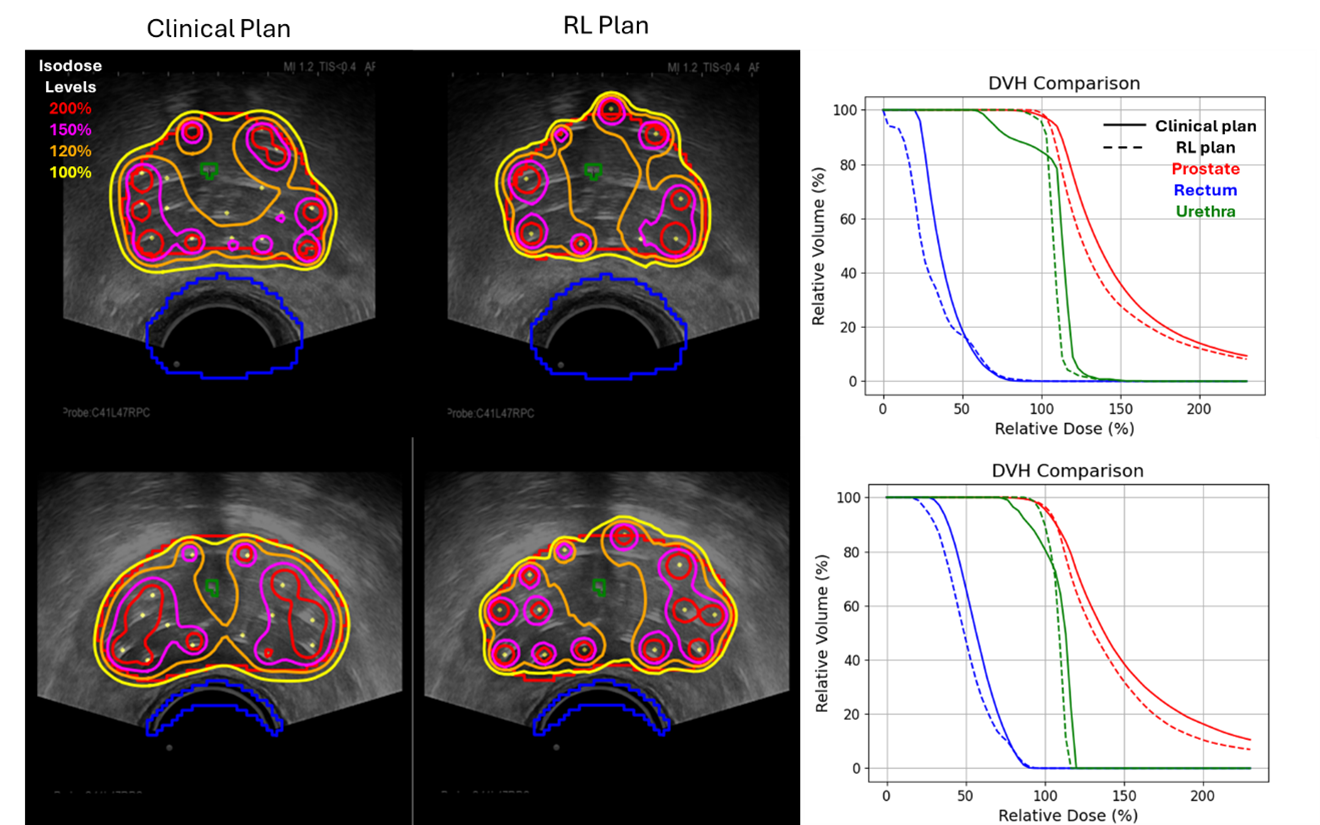}
		
		\noindent Figure 3 Isodose lines of (left) clinical plan and (middle) RL plan, and (right) DVH comparisons for two exemplary patients (upper and bottom). 
	\end{figure}
	
	Table 2 summarizes the detailed dosimetric metrics and number of needles used across all 10 evaluation patients. On average, RL plans and clinical plans achieved similar prostate coverage (Prostate V100) and rectal sparing (Rectum D2cc), with no statistically significant differences. However, the RL plans yielded significantly lower values for both Prostate V150 (indicating reduced hotspots) and Urethra D20\% (indicating improved OAR sparing), both achieving statistical significance (p < 0.05).
	
	Notably, three of the clinical plans failed to meet the clinical urethral constraint (Urethra D20\% < 120\%), whereas all RL-generated plans satisfied this requirement. Furthermore, the RL framework achieved this improved dosimetric performance while using an average of two fewer needles per patient compared to clinical plans, potentially reducing procedure time and invasiveness.
	
	\begin{table}[]
		\caption{The dosimetrics metrics and number of used needles of clinical and RL plans.}
		\label{tab:my-table}
		\resizebox{\textwidth}{!}{%
			\begin{tabular}{ccccccccccc}
				\hline
				& \multicolumn{2}{c}{Prostate V100} & \multicolumn{2}{c}{Prostate V150} & \multicolumn{2}{c}{Rectum D2cc} & \multicolumn{2}{c}{Urethra D20\%} & \multicolumn{2}{c}{\# of needles} \\ \cline{2-11} 
				& Clinical & RL & Clinical & RL & Clinical & RL & Clinical & RL & Clinical & RL \\ \cline{2-11} 
				P01 & 97.8 & 99.0 & 36.1 & 27.9 & 61.5 & 61.9 & 117.7 & 111.4 & 16 & 9 \\
				P02 & 95.7 & 96.7 & 38.7 & 32.1 & 73.8 & 68.8 & 116.8 & 112.1 & 14 & 13 \\
				P03 & 97.8 & 97.3 & 30.6 & 27.6 & 62.3 & 63.8 & 110.0 & 109.5 & 15 & 12 \\
				P04 & 95.8 & 93.6 & 32.3 & 24.5 & 74.1 & 70.1 & 114.9 & 110.2 & 12 & 13 \\
				P05 & 94.6 & 94.5 & 31.3 & 30.9 & 67.3 & 74.9 & 126.2 & 112.7 & 8 & 6 \\
				P06 & 98.8 & 98.2 & 35.9 & 24.6 & 44.6 & 46.1 & 118.9 & 110.7 & 15 & 10 \\
				P07 & 97.3 & 94.6 & 35.8 & 29.0 & 67.3 & 71.0 & 122.8 & 118.3 & 10 & 9 \\
				P08 & 97.3 & 94.3 & 34.4 & 20.9 & 70.8 & 66.0 & 122.1 & 109.7 & 8 & 12 \\
				P09 & 96.2 & 97.0 & 38.3 & 31.7 & 68.4 & 68.4 & 111.1 & 113.9 & 16 & 13 \\
				P10 & 95.2 & 95.5 & 38.8 & 26.0 & 72.6 & 65.8 & 118.7 & 111.1 & 14 & 12 \\
				P11 & 97.3 & 96.7 & 27.0 & 24.8 & 62.6 & 63.4 & 108.8 & 111.3 & 14 & 8 \\
				\begin{tabular}[c]{@{}c@{}}Mean\\    ±std\end{tabular} & \begin{tabular}[c]{@{}c@{}}96.7\\    ±1.3\end{tabular} & \begin{tabular}[c]{@{}c@{}}96.1\\    ±1.7\end{tabular} & \begin{tabular}[c]{@{}c@{}}34.5\\    ±3.8\end{tabular} & \begin{tabular}[c]{@{}c@{}}27.3\\    ±3.5\end{tabular} & \begin{tabular}[c]{@{}c@{}}65.9\\    ±8.4\end{tabular} & \begin{tabular}[c]{@{}c@{}}65.5\\    ±7.5\end{tabular} & \begin{tabular}[c]{@{}c@{}}117.1\\    ±5.5\end{tabular} & \begin{tabular}[c]{@{}c@{}}111.9\\    ±2.5\end{tabular} & \begin{tabular}[c]{@{}c@{}}13\\    ±3\end{tabular} & \begin{tabular}[c]{@{}c@{}}11\\    ±2\end{tabular} \\
				P-value & \multicolumn{2}{c}{0.216} & \multicolumn{2}{c}{\textless{}0.001} & \multicolumn{2}{c}{0.733} & \multicolumn{2}{c}{0.009} & \multicolumn{2}{c}{0.037} \\ \hline
			\end{tabular}%
		}
	\end{table}
	
	\section{Discussion}
	This study presents the first demonstration that reinforcement learning (RL) can be used to generate a complete, clinically practical prostate HDR brachytherapy treatment plan directly from patient anatomy. While further prospective evaluation is necessary to confirm its applicability across broader clinical scenarios, the approach holds significant promise—particularly for reducing the experience gap in brachytherapy planning. We proposed a novel, AI-driven workflow to automate prostate HDR treatment planning in the pre-procedural stage. The RL agent is capable of generating not only needle placements but also dwell positions and times, thereby offering physicians a reliable preplanning tool before needle insertion. This form of guidance can be instrumental in boosting procedural confidence and consistency, especially for less experienced operators.
	
	By minimizing reliance on procedural experience, the use of AI can reduce variability in implant quality, which is particularly valuable in clinics with low brachytherapy volumes. Such clinics often face challenges in maintaining procedural expertise, which may limit their ability or willingness to offer brachytherapy. The proposed method can help bridge this gap, enabling broader adoption of HDR brachytherapy without compromising treatment quality. High-volume centers may also benefit from this approach by incorporating it as a pre-procedural quality control mechanism within their clinical workflow.
	
	Our RL-based approach learns to optimize needle placement and dwell time through trial-and-error interaction with a simulation environment. Remarkably, we demonstrate that the agent can learn generalized planning strategies—such as peripheral loading—from training on a single patient. This suggests strong generalization capability and reflects underlying anatomical consistency across patients. The minimal data requirement further enhances the clinical feasibility of training and deploying this system in data-constrained settings.
	
	Previous studies have investigated AI-driven needle placement for prostate HDR brachytherapy.\cite{RN1} For example, Lei et al. developed an atlas-based method that identifies a similar case from a precompiled dataset and deforms the needle pattern using deep learning–based image registration. However, their method lacks dwell time information and requires further optimization after generating the needle configuration. Moreover, the reliance on past clinical plans as atlases inherently limits the achievable plan quality to that of historical cases. In contrast, our RL-based framework generates complete treatment plans—including dwell times—independently of existing clinical data. This independence allows the RL agent to develop new strategies that, in our study, often outperformed historical clinical plans in dosimetric quality.
	
	The RL agent operates based on anatomical contours of the prostate and OARs, which can be manually drawn by clinicians or automatically generated using AI-based segmentation tools. \cite{RN6177, RN6166, RN6175, RN6180, RN6176, RN1444, RN6137, RN6178, RN6174, RN6173, RN6179} For instance, Lei et al. proposed a multidirectional V-Net and later an anchor-free Mask CNN to segment structures on TRUS \cite{RN1444, RN6137}. Wang et al. have also demonstrated real-time segmentation of the prostate and OARs using open-source models \cite{RN6186}. With AI-based segmentation, structure contours can be generated rapidly and reviewed by clinicians, enabling efficient integration with our RL-based planner.
	
	Although the RL-generated plans generally outperform clinical plans, we note that RL does not guarantee a globally optimal solution. Instead, the agent converges toward locally optimal strategies derived from its own learning trajectory. Traditional optimization methods have attempted to solve this problem deterministically. For example, Wang et al. formulated prostate HDR planning as a convex optimization problem \cite{RN6187}. While such methods may theoretically yield globally optimal plans for a given parameter set, they require careful manual parameter tuning. In practice, finding the right parameters can be time-consuming. As shown in their paper, the solving time is 1.5 minutes, while it still requires 10 minutes to get a final plan due to the manual parameter adjustment. In other words, the traditional optimization method may give the global optimal solution for a specific group of parameters, while the choice of these parameters is another open question.
	
	There are several limitations to this study. First, the evaluation was performed on a relatively small patient cohort from a single institution. In future work, we plan to expand the dataset to more comprehensively evaluate the method’s performance across diverse anatomies and under varying training conditions, including alternative policy architectures and hyperparameter configurations. Furthermore, this was a retrospective study; however, the promising results support future prospective studies and multi-institutional collaborations to validate the generalizability of our approach in real-world clinical settings.
	
	An important future direction involves extending our framework to support adaptive intraoperative planning. In clinical practice, needle placements often deviate from preplanned positions due to operator variability or anatomical changes during the procedure. Integrating real-time feedback—such as intraoperative contours and updated needle positions—into the RL framework would allow the model to dynamically update the plan as needles are placed \cite{RN6182, RN6181, RN6185, RN6184, RN6183, RN6077, RN6076, RN6075}. This would represent a significant advancement toward intelligent, closed-loop adaptive brachytherapy.

	\section{Conclusion}
	In this study, we introduced a novel reinforcement learning (RL) framework for automated treatment planning in prostate high-dose-rate (HDR) brachytherapy. Our method generates complete treatment plans—including needle positions and dwell times—directly from patient anatomical contours, without relying on prior clinical plans. The RL-generated plans achieved comparable or superior dosimetric quality to clinical plans while using fewer needles, demonstrating improved prostate dose uniformity and better urethral sparing. Importantly, the model required minimal training data and exhibited strong generalizability across patients, highlighting its potential for broad clinical applicability.

	This work represents a significant step toward intelligent, preoperative planning for HDR brachytherapy and lays the foundation for future development of adaptive, intraoperative planning systems. With further validation and integration, the proposed approach could streamline clinical workflows, reduce reliance on procedural experience, and enhance treatment consistency and quality across institutions.

	\noindent 
	
	\bibliographystyle{plainnat}  
	\bibliography{arxiv}      
	
\end{document}